\documentclass[preprint,authoryear,12pt]{elsarticle}%

\usepackage{amssymb}
\usepackage{graphicx}
\usepackage{epsfig}
\usepackage{amsthm}
\usepackage{longtable}
\usepackage{lscape}
\usepackage{amsmath}

\journal{Journal of High Energy Astrophysics}

\begin{document}

\begin{frontmatter}



\title{Long GRBs as a Tool to Investigate Star Formation in Dark Matter Halos}


\author[label1]{Jun-Jie Wei}
\author[label2,label3]{Jing-Meng Hao}
\author[label1,label4]{Xue-Feng Wu\corref{dip}}
\ead{xfwu@pmo.ac.cn}
\author[label5]{Ye-Fei Yuan}
\ead{yfyuan@ustc.edu.cn}
\address[label1]{Purple Mountain Observatory, Chinese Academy of Sciences, Nanjing 210008, China}
\def\astrobj#1{#1}
\address[label2]{Center for Astrophysics, Guangzhou University, Guangzhou 510006, China}
\address[label3]{Astronomy Science and Technology Research Laboratory of Department of Education of Guangdong Province, Guangzhou 510006, China}
\address[label4]{Joint Center for Particle, Nuclear Physics and Cosmology, Nanjing
University-Purple Mountain Observatory, Nanjing 210008, China}
\address[label5]{Key Laboratory for Research in Galaxies and Cosmology CAS, Department of Astronomy,
University of Science and Technology of China, Hefei, Anhui 230026, China}
\cortext[dip]{Corresponding author.}

\begin{abstract}
First stars can only form in structures that are suitably dense, which can be parametrized
by the minimum dark matter halo mass $M_{\rm min}$. $M_{\rm min}$ must plays an important
role in star formation. The connection of long gamma-ray bursts
(LGRBs) with the collapse of massive stars has provided a good opportunity for probing
star formation in dark matter halos. We place some constraints on $M_{\rm min}$
using the latest $Swift$ LGRB data. We conservatively consider that LGRB rate is proportional
to the cosmic star formation rate (CSFR) and an additional evolution parametrized as
$(1+z)^{\alpha}$, where the CSFR model as a function of $M_{\rm min}$. Using the
$\chi^{2}$ statistic, the contour constraints on the $M_{\rm min}$--$\alpha$ plane show that
at the $1\sigma$ confidence level, we have $M_{\rm min}<10^{10.5}$ $\rm M_{\odot}$ from
118 LGRBs with redshift $z<4$ and luminosity $L_{\rm iso}>1.8\times10^{51}$ erg $\rm s^{-1}$.
We also find that adding 12 high-\emph{z} $(4<z<5)$ LGRBs (consisting of 104 LGRBs with $z<5$
and $L_{\rm iso}>3.1\times10^{51}$ erg $\rm s^{-1}$) could result in much tighter
constraints on $M_{\rm min}$, for which, $10^{7.7}\rm M_{\odot}<M_{\rm min}<10^{11.6}\rm M_{\odot}$ ($1\sigma$).
Through Monte Carlo simulations, we estimate that future five years
of Sino-French spacebased multiband astronomical variable objects monitor (\emph{SVOM}) observations
would tighten these constraints to $10^{9.7}\rm M_{\odot}<M_{\rm min}<10^{11.3}\rm M_{\odot}$.
The strong constraints on $M_{\rm min}$ indicate that LGRBs are a new promising tool
for investigating star formation in dark matter halos.
\end{abstract}


\end{frontmatter}


\def\astrobj#1{#1}
\section{Introduction}
\label{sect:intro}

Gamma-ray bursts (GRBs) are the most luminous explosive events in the cosmos,
which can be detected even out to the edge of the Universe. To date, the highest redshift
of GRBs is $\sim9.4$ \citep[GRB 090429B;][]{Cucchiara11}. So GRBs are considered as a
powerful tool to probe the properties of the high-\emph{z} Universe, including high-\emph{z}
star formation history \citep[e.g.,][]{Chary07,Yuksel08,Kistler09,Trenti12,Wei14}, metal enrichment history \citep{Wang12},
and the dark matter particle mass \citep{de Souza13}. Theoretically,
it is widely accepted that long bursts (LGRBs) with durations $T_{\rm 90}>2$ s
\citep[where $T_{\rm 90}$ is the interval observed to contain $90\%$ of the prompt emission;][]{Kouveliotou93}
are powered by the core collapse of massive stars \citep[e.g.,][]{Woosley93,Paczynski98,Woosley06},
which have been strongly supported by several confirmed associations
between LGRBs and Type Ic supernovae \citep{Stanek03,Hjorth03,Chornock10}.
The collapsar model suggests that the cosmic GRB rate should in principle trace
the cosmic star formation rate \citep[CSFR;][]{Totani97,Wijers98,Blain00,Lamb00,Porciani01,
Piran04,Zhang04,Zhang07}.

Thanks to the great contribution of the $Swift$ satellite \citep{Gehrels04},
the number of GRBs with measured redshifts has increased rapidly over the last decade.
Surprisingly, the $Swift$ data seems to indicate that the rate of LGRBs does not strictly
trace the CSFR, but instead implying some kind of additional evolution
\citep{Daigne06a,Guetta07,Le07,Salvaterra07,Kistler08,Kistler09,Li08,Salvaterra09,Salvaterra12,Campisi10,
Qin10,Wanderman10,Cao11,Virgili11,Elliott12,Lu12,Robertson12,Wang13,Wei14}. The observed discrepancy between the LGRB rate and the CSFR
is used to be described by an enhanced evolution parametrized as $(1+z)^{\alpha}$ \citep[e.g.,][]{Kistler08}.
Many mechanisms have been proposed to explain the enhancement, such as cosmic metallicity evolution
\citep{Langer06,Li08}, an evolution in the stellar initial mass function
\citep{Xu09,Wang11}, and an evolution in the GRB luminosity function \citep{Virgili11,Salvaterra12,Tan13,Tan15}.

However, it should be emphasized that the prediction on the LGRB rate strongly relates to the star
formation rate models. With different star formation history models, the results on the discrepancy
between LGRB rate and CSFR could change in some degree \citep[see][]{Virgili11,Hao13}.
There are many forms of CSFR available in the literature. Most previous studies
\citep[e.g.,][]{Kistler08,Kistler09,Li08,Salvaterra09,Salvaterra12,Robertson12,Wei14} adopted
the widely accepted CSFR model of \citet{Hopkins06}, which provides a good piecewise-linear fit to
the ultraviolet and far-infrared observations. But, it is obviously that the empirical fit will
vary depending on both the functional form and the observational data used. \citet{Hao13} confirmed that
LGRBs were still biased tracers of the CSFR model derived from the empirical fit of \citet{Hopkins06}. While, using
the self-consistent CSFR model calculated from the hierarchical structure formation scenario, they found that
large number of LGRBs occur in dark matter halos with mass down to $10^{8.5}M_{\odot}$ could give an
alternative explanation for the CSFR--LGRB rate discrepancy.

The fact that stars can only form in structures that are suitably dense, which can be parametrized
by the minimum mass $M_{\rm min}$ of a dark matter halo of the collapsed structures where
star formation occurs. Structures with masses smaller than $M_{\rm min}$ are considered as part of
the intergalactic medium and do not take part in the star formation process. Thus, the minimum halo mass
$M_{\rm min}$ must plays an important role in star formation. Some observational data
have been used to constrain $M_{\rm min}$ in several instance, including the following
representative cases: \citet{Daigne06b} showed that with a minimum halo mass of $10^{7}-10^{8}M_{\odot}$
and a moderate outflow efficiency, they were able to reproduce both the current baryon fraction and
the early chemical enrichment of the intergalactic medium; \citet{Bouche10} found that
a minimum halo mass $M_{\rm min}\simeq10^{11}M_{\odot}$ was required in their model to
simultaneously account for the observed slopes of the star formation rate--mass and Tully--Fisher relations;
\citet{Munoz11} found that the observed galaxy luminosity function was best fit with
a minimum halo mass per galaxy of $10^{9.4^{+0.3}_{-0.9}}M_{\odot}$.

The collapsar model suggests that LGRBs constitute an ideal tool to investigate star formation
in dark matter halos. The expected GRB redshift distributions can be calculated from the self-consistent CSFR model
as a function of the minimum halo mass $M_{\rm min}$. Thus, $M_{\rm min}$ can be constrained by directly comparing
the observed and expected redshift distributions. In this paper, we extend the work of \citet{Hao13}
by presenting robust limits on $M_{\rm min}$ using the latest $Swift$ GRB data. Since the latest data
have many redshift measurements, a reliable statistical analysis is now possible. This analysis
not only provides a better understanding of the high-\emph{z} CSFR using the LGRB data, but also indicates
that LGRBs can be a new tool to constrain the minimum halo mass. The outline of this paper is as follows.
In Section~2, we will briefly describe the star formation model we have adopted and demonstrate the impact of
the minimum halo mass $M_{\rm min}$ on the CSFR. In Section~3, we will present the method for calculating
the theoretical GRB redshift distribution, and then in Section~4 show direct constraints on the numerical value
of $M_{\rm min}$ from the latest $Swift$ GRB data. In Section~5, we will discuss possible future constraints
using a mock sample. Finally, we will end with our conclusions in Section~6.

Throughout we use the cosmological parameters from the \emph{Wilkinson Microwave
Anisotropy Probe} (WMAP) nine-year data release \citep{Hinshaw13}, namely
$\Omega_{\rm m}=0.286$, $\Omega_{\Lambda}=0.714$, $\Omega_{\rm b}=0.0463$,
$\sigma_{8}=0.82$ and $h=0.69$.
\section{The cosmic star formation}
\label{sect:CSFR}
In the framework of hierarchical structure formation, the self-consistent CSFR model can be obtained
by solving the evolution equation of the total gas density that takes into account
the baryon accretion rate, the ejection of gas by stars, and the stars formed
through the transfer of baryons in the dark matter halos \citep[see][]{Pereira10}.
The baryon accretion rate stands for the process of structure formation,
which governs the size of the reservoir of baryons available for star formation \citep{Daigne06b}.
In this section, we will briefly summarize how to obtain the CSFR from the hierarchical model,
which is developed by \citet{Pereira10}.

In the hierarchical formation scenario, the comoving abundance of collapsed dark
matter halos can be determined based on the Press--Schechter (P--S) like formalism \citep{Press74}.
We adopt the most popularly used halo mass function, named the Sheth--Tormen mass
function \citep{Sheth99}, which is similar to the form of the P--S mass function:
\begin{equation}
f_{\rm ST}(\sigma)=A\sqrt{\frac{2a_{1}}{\pi}}\left[1+\left(\frac{\sigma^{2}}{a_{1}\delta^{2}_{c}}\right)^{p}\right]
\frac{\delta_{c}}{\sigma}\exp\left(-\frac{a_{1}\delta^{2}_{c}}{2\sigma^{2}}\right)\;,
\end{equation}
where the parameter $\delta_{c}=1.686$ could be explained physically as the
linearly extrapolated overdensity of a top-hat spherical density perturbation
at the time of maximum compression. The choice of values $A=0.3222$, $a_{1}=0.707$,
and $p=0.3$ gives the best fit to mass functions derived from numerical
simulations over a broad range of redshifts and masses.
The number density of dark matter halos with mass $M$, $n_{\rm ST}(\emph{M},\emph{z})$,
can be related to $f_{\rm ST}(\sigma)$ by
\begin{equation}
\frac{\rm d\emph{n}_{\rm ST}(\emph{M},\emph{z})}{\rm d\emph{M}}=\frac{\rho_{\rm m}}{M}\frac{\rm d \ln \sigma^{-1}}{\rm d\emph{M}}f_{\rm ST}(\sigma)\;,
\end{equation}
where $\rho_{\rm m}$ is the mean mass density of the Universe.
The variance of the linearly density field $\sigma(M,z)$ is given by
\begin{equation}
\sigma^{2}(M,z)=\frac{D^{2}(z)}{2\pi^{2}}\int^{\infty}_{0}k^{2}P(k)W^{2}(k,M) \rm d\emph{k}\;,
\end{equation}
where the primordial power spectrum $P(k)$ is smoothed with a real space top-hat
filter function $W(k,M)$, $D(z)$ is the growth factor of linear perturbations normalized to
$D=1$ at the present epoch and the redshift dependence enters only through $D(z)$.

The baryon distribution is considered to be tracing the dark matter
distribution without any bias, which means the baryons density is completely
proportional to the density of dark matter. Note that first stars can form only in
structures that are suitably dense, which can be parametrized by the minimum
dark matter halo mass $M_{\rm min}$. Thus, star formation will be suppressed
when the halo mass below $M_{\rm min}$. In fact, the suppression in star
formation is time dependent, i.e., the minimum mass $M_{\rm min}$ should evolve with
\emph{z} as the cooling processes of the hot gas in structures depend on the
chemical composition and ionizing state of the gas \citep[see][]{Daigne06b}. However,
the process of evolution is very complex. It is beyond the scope of this study
to consider the detailed analysis on evolution, so we would like to keep $M_{\rm min}$ as a constant
and set it as a free parameter in this model, as those authors did in their works
\citep[see, e.g.,][]{Daigne06b,Pereira10,Munoz11}.
Therefore, the fraction of baryons inside collapsed halos at redshift \emph{z}
is given by
\begin{equation}
f_{\rm b}(z)=\frac{\int^{\infty}_{M_{\rm min}}n_{\rm ST}(M, z)M \;\rm d\emph{M}}
{\int^{\infty}_{0}n_{\rm ST}(M, z)M \;\rm d\emph{M}} \;.
\end{equation}
With the fraction, the baryons accretion rate $a_{\rm b}(t)$ as a function of redshift,
which accounts for the formation of structures, can be calculated by
\begin{equation}
a_{\rm b}(t)=\Omega_{\rm b}\rho_{\rm c}\left(\frac{\rm d\emph{t}}{\rm d\emph{z}}\right)^{-1}
\left|\frac{\rm d \emph{f}_{b}(\emph{z})}{\rm d\emph{z}}\right| \;,
\end{equation}
where $\rho_{\rm c}=3H^{2}_{0}/8\pi G$ is the critical density of the Universe.

For ease of calculation, we consider the star formation rate satisfying the Schmidt law \citep{Schmidt59,Schmidt63},
as \citet{Pereira10} did in their treatment. The Schmidt law suggests that the star formation rate $\dot{\rho}_{\star}(t)$
is directly proportional to the local gas density $\rho_{\rm g}(t)$, which can be simply expressed as
\begin{equation}
\frac{\rm d^{2} \emph{M}_{\star}}{\rm d\emph{V} \; d\emph{t}}=
\dot{\rho}_{\star}(t)=k\rho_{\rm g}(t)\;,
\end{equation}
where $k$ is a constant.

The mass ejected from stars, which is returned to the interstellar medium through
winds and supernovae, is given by
\begin{equation}
\frac{\rm d^{2} \emph{M}_{\rm ej}}{\rm d\emph{V} \; d\emph{t}}=
\int^{m_{\rm sup}}_{m(t)}(m-m_{\rm r})\Phi(m)\dot{\rho}_{\star}(t-\tau_{m})\; \rm d\emph{m} \;,
\end{equation}
where $m(t)$ corresponds to the stellar mass whose lifetime is equal to $t$. The mass of the remnant,
$m_{\rm r}$, depends on the progenitor mass \citep[see][]{Pereira10}. The stellar initial mass function
$\Phi(m)$ follows the \citet{Salpeter55} form, $\Phi(m)=Am^{-2.35}$, with the mass range
$[m_{\rm inf}, m_{\rm sup}]$, where $m_{\rm inf}=0.1 \;\rm M_{\odot}$ and $m_{\rm sup}=140
\;\rm M_{\odot}$. $\tau_{m}$ is the lifetime of a star with mass $m$, which is calculated
using the metallicity-independent fit of \citet{Scalo86} and \citet{Copi97}.

Combining Equations~(5), (6) and (7), the evolution of the total gas density
($\rho_{\rm g}$) that determines the star formation history in the dark matter halos
can be written down as
\begin{equation}
\dot{\rho}_{\rm g}=-\frac{\rm d^{2} \emph{M}_{\star}}{\rm d\emph{V} \; d\emph{t}}
+\frac{\rm d^{2} \emph{M}_{\rm ej}}{\rm d\emph{V} \; d\emph{t}}+a_{\rm b}(t)\;.
\end{equation}
Finally, we can produce the function $\rho_{\rm g}(t)$ at each time $t$ (or redshift $z$)
by solving Equation~(8). Once obtained $\rho_{\rm g}(t)$, we can calculate the CSFR
$\dot{\rho}_{\star}(t)$ according to Equation~(6), i.e., $\dot{\rho}_{\star}=k\rho_{\rm g}$,
where the constant $k=1/\tau_{\rm s}$ denotes the inverse of the timescale for star
formation. Consistent with previous works \citep[see, e.g.,][]{Pereira10,Hao13},
we use $\tau_{\rm s}=2.0$ Gyr as the timescale of star formation and consider that
the star formation starts at redshift $z_{\rm ini}=20$. The CSFR is normalized to
$\dot{\rho}_{\star}=0.016$ $\rm M_{\odot}$ $\rm yr^{-1}$ $\rm Mpc^{-3}$ at $z=0$ \citep{Hopkins04,Hopkins07}.

In Fig.~1, we show the CSFR obtained from the self-consistency models as a function of the minimum
halo mass $M_{\rm min}$ (see Equation~4). The observational CSFR taken from \citet{Hopkins04,Hopkins07}
and \citet{Li08}, which are based on the observations of other authors who
are listed in these publications, are also shown for comparison. One can see from this plot that $\dot{\rho}_{\star}(z)$
is very sensitive to the minimum mass $M_{\rm min}$, especially at high-\emph{z}.
In addition, all of these models have good agreement with observational data at $z\leq6$.

\begin{figure}[hp]
\vskip-0.3in
\centerline{\includegraphics[angle=0,scale=0.6]{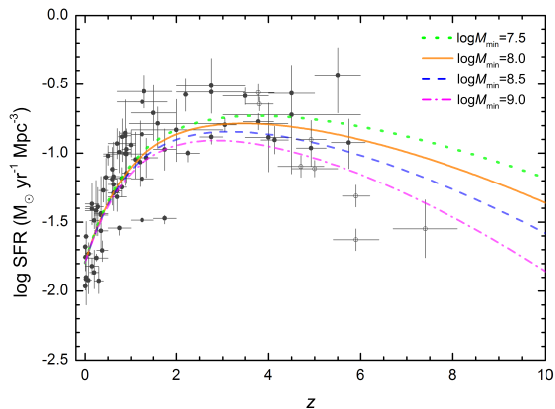}}
\vskip-0.2in
\caption{The CSFR $\dot{\rho}_{\star}(z)$ derived from the self-consistency models compared to the observational data
taken from \citet{Hopkins04,Hopkins07} (dots) and \citet{Li08} (circles). The curves represent models with
a minimum halo mass $M_{\rm min}=10^{7.5}$ $\rm M_{\odot}$, $10^{8.0}$ $\rm M_{\odot}$,
$10^{8.5}$ $\rm M_{\odot}$, and $10^{9.0}$ $\rm M_{\odot}$, respectively.}
\end{figure}
\section{The GRB technique}
\label{sect:tech}
As discussed above, we assume that the relationship between the comoving GRB rate
and the CSFR density $\dot{\rho}_{\star}$ can be expressed as $\dot{n}_{\rm GRB}(z)
=\varepsilon(z)\dot{\rho}_{\star}(z)$, where $\varepsilon(z)$ accounts for the
formation efficiency of LGRBs. Note that the CSFR
can be obtained from the self-consistency model with a free parameter $M_{\rm min}$
as described in Section~2. The expected redshift distribution of GRBs is given as
\begin{equation}
\frac{\rm d\emph{N}}{\rm d\emph{z}}=F(z)\frac{\varepsilon(z)\dot{\rho}_{\star}(z)}{\langle f_{\rm beam} \rangle}
\frac{\rm d \emph{V}_{com}/d\emph{z}}{1+z} \;,
\end{equation}
where $F(z)$ represents the ability both to detect the GRB and to obtain the redshift,
$\langle f_{\rm beam} \rangle$ is the beaming factor, the factor $(1+z)^{-1}$ accounts
for the cosmological time dilation, and $\rm d \emph{V}_{com}/d\emph{z}$ is the comoving
volume element. As discussed in detail in \citet{Kistler08}, $F(z)$ can be treated
as a constant ($F_{0}$) when we only consider the bright bursts with luminosities sufficient
to be observed within an entire redshift range.

There is a general agreement about the fact that the LGRB rate does not strictly
follow the CSFR but is actually enhanced by some unknown mechanisms at high-\emph{z}.
Several evolution scenarios have been considered to explain the observed enhancement,
including the GRB rate density evolution \citep{Kistler08,Kistler09}, cosmic metallicity
evolution \citep{Langer06,Li08}, stellar initial mass function evolution \citep{Xu09,Wang11},
and luminosity function evolution \citep{Virgili11,Salvaterra12,Tan13,Tan15}.
In a word, there are much debate in the mechanisms responsible for the enhancement.
For simplicity, we adopt the density evolution model and parametrize the evolution in the
GRB rate as $\varepsilon(z)=\varepsilon_{0}(1+z)^{\alpha}$, where $\varepsilon_{0}$ is a constant
that includes the fraction of stars that produce long GRBs.
Here, we conservatively keep $\alpha$ as a free parameter. So there are
two free parameters $M_{\rm min}$ and $\alpha$ in our calculation.

The expected number of GRBs within a redshift range $z_{1}\leq z \leq z_{2}$, for each
combination $\textbf{P}\equiv\{M_{\rm min}, \alpha\}$, can be described as
\begin{equation}
\begin{split}
N(z_{1}, z_{2}; \textbf{P})=\Delta t \frac{\Delta \Omega}{4\pi}\int^{z_{2}}_{z_{1}}F(z)\varepsilon(z)
\frac{\dot{\rho}_{\star}(z; M_{\rm min})}{\langle f_{\rm beam} \rangle}\frac{\rm d \emph{V}_{com}/d\emph{z}}{1+z}\; \rm d\emph{z}\\
=\mathcal{A}\int^{\emph{z}_{2}}_{\emph{z}_{1}}(1+\emph{z})^{\alpha}\dot{\rho}_{\star}(\emph{z}; M_{\rm min})\frac{\rm d \emph{V}_{com}/d\emph{z}}{1+\emph{z}}\; \rm d\emph{z}\;,
\end{split}
\end{equation}
where the constant $\mathcal{A}=\Delta t \Delta \Omega F_{0}\varepsilon_{0}/4\pi\langle f_{\rm beam} \rangle$
depends on the total observed time, $\Delta t$, and the angular sky coverage, $\Delta \Omega$.
In order to remove the dependence on $\mathcal{A}$, we can simply construct the cumulative
redshift distribution of GRBs over the redshift range $0 < z < z_{\rm max}$, normalized to $N(0, z_{\rm max})$,
as
\begin{equation}
N(<z|z_{\rm max})=\frac{N(0, z)}{N(0, z_{\rm max})} \;.
\end{equation}
\section{Constraints from \emph{Swift} long GRBs}
\label{sect:constraints}
Our LGRB sample is taken from \citet{Wei14}, which is consisted of long GRBs detected by $Swift$
up to 2013 July. Most of the data are collected from the samples presented in \citet{Butler07,Butler10}
and \citet{Sakamoto11}. Redshift measurements are strongly biased towards optically bright
afterglows, and are more easily made when the afterglow is not obscured by dust \citep[see e.g.][]{Greiner11}.
The phenomenon of so-called dark GRBs with suppressed optical counterparts could
influence whether the observed cumulative redshift distribution $N(<z)$ is representative of that
for all long GRBs. Therefore, it is important to add the redshift of dark GRBs. \citet{Wei14}
also included dark GRBs from \citet{Perley09}, \citet{Greiner11}, \citet{Kruhler11},
\citet{Hjorth12}, and \citet{Perley13}.\footnote{Several of the dark bursts with redshift
upper limits are not included in this work.} With the information of redshift $z$,
burst duration $T_{90}$, and isotropic-equivalent energy $E_{\rm iso}$ for each GRB taken from
\citet{Wei14}, we calculate the isotropic-equivalent luminosities using $L_{\rm iso}=E_{\rm iso}/
[T_{90}/(1+z)]$.

\begin{figure}[hp]
\vskip-0.2in
\centerline{\includegraphics[angle=0,scale=0.6]{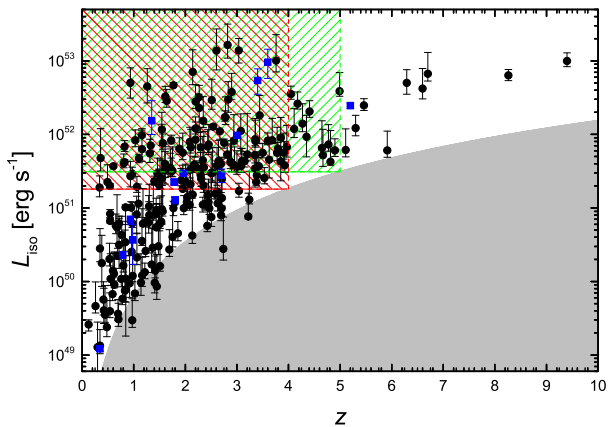}}
\vskip-0.2in
\caption{The luminosity-redshift distribution of 244 \emph{Swift} GRBs from the catalog of \citet{Wei14}.
The blue squares represent 13 dark bursts with firm redshift determinations.
The gray shaded region approximates the \emph{Swift} detection threshold.
The red box corresponds to 118 GRBs with $z<4$ and $L_{\rm iso}>1.8\times10^{51}$ erg $\rm s^{-1}$,
and the green box corresponds to 104 GRBs with $z<5$ and $L_{\rm iso}>3.1\times10^{51}$ erg $\rm s^{-1}$.}
\end{figure}

Our final sample includes 244 GRBs with firm redshift determinations, whose luminosity-redshift
distribution is shown in Fig.~2. The shaded region represents the effective
detection threshold of \emph{Swift}. The luminosity threshold can be approximated using a bolometric
energy flux limit $F_{\rm lim}=1.2\times10^{-8}$ erg $\rm cm^{-2}$ $\rm s^{-1}$ \citep{Kistler08},
i.e., $L_{\rm lim}=4\pi D_{L}^{2}F_{\rm lim}$,
where $D_{L}$ is the luminosity distance. The sensitivity of \emph{Swift}/Burst Alert Telescope (BAT)
is very difficult to parametrize exactly \citep{Band06}. In order to avoid the influence of \emph{Swift}
threshold, we will adopt a model-independent approach by selecting only GRBs with $L_{\rm iso}>L_{\rm lim}$
and $z<4$, as \citet{Kistler08} did in their treatment. The cut in luminosity is chosen to be equal to the
threshold at the highest redshift of the sample, i.e., $L_{\rm lim}(z=4)\approx1.8\times10^{51}$ erg $\rm s^{-1}$.
The cut in luminosity and redshift can reduce the selection effects by removing many low-$z$, low-$L_{\rm iso}$
bursts that could not have been observed at higher redshift. With these conditions, we have 118 GRBs
in this sub-sample. These data are delimited by the red shaded region in Fig.~2.

\begin{figure}[hp]
\vskip-0.2in
\centerline{\includegraphics[angle=0,scale=1.0]{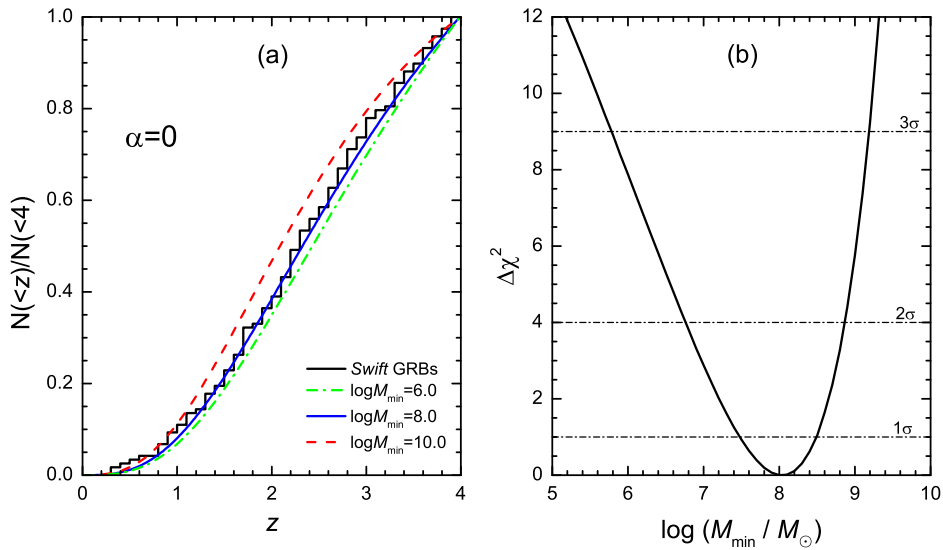}}
\vskip-0.2in
\caption{(a): cumulative redshift distribution of 118 \emph{Swift} GRBs with $z<4$
    and $L_{\rm iso}>1.8\times10^{51}$ erg $s^{-1}$ (steps). The expected redshift
    distributions inferred from the self-consistent star formation rate model
    for $\alpha=0$ as a function of the minimum halo mass $M_{\rm min}$ are also shown (from bottom to top):
    the green dot-dashed line corresponds to $M_{\rm min}=10^{6.0}$ $\rm M_{\odot}$,
    the blue solid line corresponds to $M_{\rm min}=10^{8.0}$ $\rm M_{\odot}$, and
    the red dashed line corresponds to $M_{\rm min}=10^{10.0}$ $\rm M_{\odot}$.
    (b): constraints on the minimum halo mass, $M_{\rm min}$, for the $\alpha=0$ model.}
\end{figure}

Fig.~3(a) shows the cumulative redshift distribution of these 118 GRBs (steps), as well as the
expected redshift distributions inferred from the self-consistent CSFR model (curves).
To evaluate the consistency between the observed and the expected GRB redshift distributions, we
make use of the one-sample Kolmogorov-Smirnov (K-S) test. In Fig.~3(a), we firstly consider the
non-evolution case (i.e., the evolutionary index $\alpha=0$), and compare the observed GRB cumulative redshift distribution
with the expected distribution for different values of $M_{\rm min}$. We can find that
the expectations from the models with a minimum halo mass $M_{\rm min}=10^{6.0}$ $\rm M_{\odot}$
(green dot-dashed line)or $M_{\rm min}=10^{10.0}$ $\rm M_{\odot}$ (red dashed line) are incompatible with the
observations. The test statistics and probability for the relevant models are presented in Table~1.
While, the model with $M_{\rm min}=10^{8.0}$ $\rm M_{\odot}$ (blue solid line)
can reproduce the observed data very well, with a maximum K-S probability of $P=0.993$, which is consistent with that of
\citet{Hao13}. This result implies that most of LGRBs occur in small dark matter halos down
to $10^{8.0}$ $\rm M_{\odot}$ can provide an alternative explanation for the discrepancy
between the LGRB rate and the CSFR, without considering the extra evolution effect (i.e., $\alpha=0$).

In order to find the best-fit parameters together with their $1\sigma$ (or $2\sigma$) confidence level,
we also optimize the model fits by minimizing the $\chi^{2}$ statistic
\begin{equation}
\chi^{2}=\sum_{i}^{n}\frac{\left[N^{\rm th}(<z_{i}|z_{\rm max})-N^{\rm obs}(<z_{i}|z_{\rm max})\right]^{2}}{\sigma_{i}^{2}}\;,
\end{equation}
where $n$ is the number of $z$ bins, $N^{\rm th}(<z_{i}|z_{\rm max})$ and $N^{\rm obs}(<z_{i}|z_{\rm max})$
are the expected and the observed (normalized) cumulative numbers of LGRBs in bin $i$, respectively.
For the observed number $N_{i}(<z_{i})$ in bin $i$, the statistical error of $N_{i}(<z_{i})$ is usually
considered to be the Poisson error, i.e., $\bar{\sigma_{i}}=\sqrt{N_{i}(<z_{i})}$, which corresponds to
the $68\%$ Poisson confidence intervals for the binned events. Since the observed cumulative number
is normalized to $N(0, z_{\rm max})$ (see Equation~11), the standard deviation errors turn to be
$\sigma_{i}=\sqrt{N_{i}(<z_{i})}/N(0, z_{\rm max})$. If the accumulated distribution is treated as
a sum of independent measurements in the different 40 $z$ bins of width $\Delta z=0.1$ between
$z=0$ and $z=4$, the results of fitting the 40 $z$ bins with different $M_{\rm min}$ are shown
in Fig.~3(b) (solid line). We see here that the best fit corresponds to $\log M_{\rm min}=
8.0_{-0.5}^{+0.5}(1\sigma)_{-1.2}^{+0.9}(2\sigma)$. It is interesting to note that \citet{Munoz11}
found the minimum halo mass capable of hosting galaxies can be around $2.5\times10^{9}$
$\rm M_{\odot}$ by fitting the observed galaxies luminosity function, in agreement with
the minimum halo mass we derive here using GRB data. With $40-1=39$ degrees of freedom,
the reduced $\chi^{2}$ for the CSFR model with an optimized minimum halo mass is $\chi_{\rm dof}^{2}=4.98/39=0.13$.
Note that taking different values for $\Delta z$ has very little impact on the best-fit results.

Next, we fix $M_{\rm min}=10^{13.0}$ $\rm M_{\odot}$ and keep $\alpha$ as a free parameter.
The theoretical GRB cumulative redshift distributions for different values of $\alpha$ are shown in Fig.~4(a).
The high halo mass (i.e., $M_{\rm min}=10^{13.0}$ $\rm M_{\odot}$) means that the suppression
of dark matter halo abundances in this model is very strong, which leads to an unrealistically
high value of $\alpha\sim5.32$ required to be roughly consistent with the observations (blue solid line),
with a K-S probability of $P=0.159$. Using the $\chi^{2}$ statistic, the constraints on $\alpha$
are shown in Fig.~4(b). For this fit, we obtain $\alpha=5.32_{-0.17}^{+0.17}(1\sigma)_{-0.32}^{+0.34}(2\sigma)$.
With $40-1=39$ degrees of freedom, the reduced $\chi^{2}$ is $\chi_{\rm dof}^{2}=39.58/39=1.01$.
However, such a high value can be ruled out by low-\emph{z} observations, which imply $\alpha\leq1.0$
\citep[e.g.,][]{Kistler09,Robertson12,Trenti12,Wei14}.
Moreover, \citet{Trenti12} suggested that there is significant star formation in faint galaxies,
it is not possible that the halo mass capable of hosting galaxies can come to be around $10^{13.0}$ $\rm M_{\odot}$.

\begin{figure}[hp]
\vskip-0.2in
\centerline{\includegraphics[angle=0,scale=1.0]{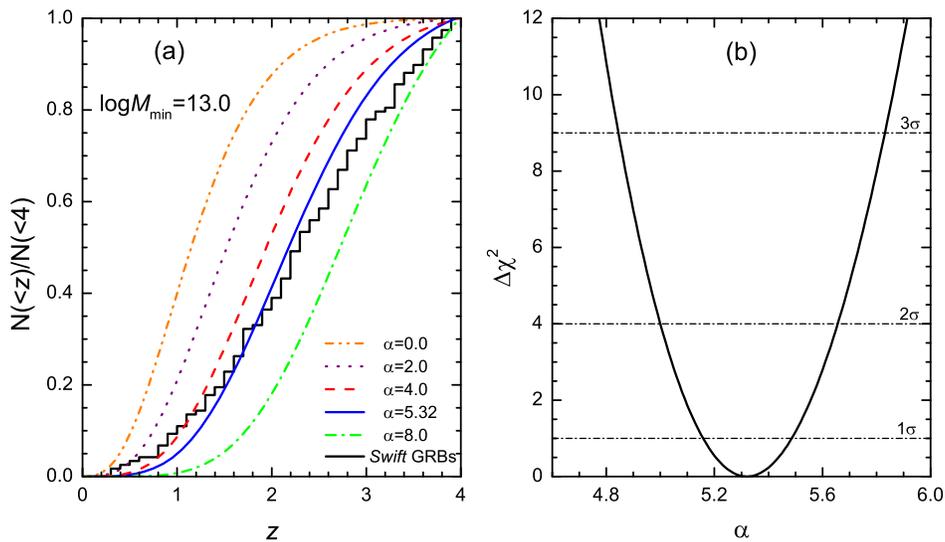}}
\vskip-0.2in
\caption{(a): same as Figure~3(a), but the expected redshift distributions are calculated
    for $M_{\rm min}=10^{13.0}$ $\rm M_{\odot}$ while different values of $\alpha$.
    From top to bottom, the orange dot-dot-dashed line represents
    $\alpha=0.0$, purple dot line $\alpha=2.0$, red dashed line $\alpha=4.0$,
    blue solid line $\alpha=5.9$, and green dot-dashed line $\alpha=8.0$.
    (b): constraints on the evolutionary index, $\alpha$, for the
    $M_{\rm min}=10^{13.0}$ $\rm M_{\odot}$ model.}
\end{figure}

\begin{center}
\begin{table}
\caption{Statistical tests of the relevant models}\label{1}
\begin{tabular}{ccc}
\hline
$\alpha$&$\log M_{\rm min}$& K-S test \\
        &$\rm (M_{\odot})$ &D-stat,~Prob \\
\hline
0.0   &  6.0   & 0.0629,~0.7257	       \\
0.0   &  8.0   & 0.0392,~0.9925	       \\
0.0   &  10.0  & 0.1235,~0.0501	       \\
0.0   &  13.0  & 0.5232,~0.0000	       \\
2.0   &  13.0  & 0.3851,~0.0000	       \\
4.0   &  13.0  & 0.1927,~0.0003	       \\
5.32   &  13.0  & 0.1024,~0.1590	   \\
8.0   &  13.0  & 0.2136,~0.0000	       \\
 \hline
\end{tabular}
\end{table}
\end{center}

If we relax the priors, and allow both $M_{\rm min}$ and $\alpha$ to be free parameters,
we can construct confidence limits in the two-dimensional parameter space ($M_{\rm min}$, $\alpha$)
by fitting the cumulative redshift distribution of 118 \emph{Swift} GRBs with $z<4$ and
$L_{\rm iso}>1.8\times10^{51}$ erg $\rm s^{-1}$, using the $\chi^{2}$ statistic.
Fig.~5(a) shows the $1\sigma-3\sigma$ constraint contours of the probability in the
($M_{\rm min}$, $\alpha$) plane. These contours show that at the $1\sigma$ level,
$-0.54<\alpha<0.99$, while $M_{\rm min}$ is weakly constrained;
only an upper limit of $10^{10.5}$ $\rm M_{\odot}$ can be set at this confidence level.
The cross indicates the best-fit pair $(\log M_{\rm min},\;\alpha)=(7.2,\;-0.15)$.

As shown in Fig.~1, the CSFR $\dot{\rho}_{\star}(z)$ is very sensitive to the
minimum mass $M_{\rm min}$, especially at high-\emph{z}. To explore the
dependence of our results on a possible bias in the high-\emph{z} bursts, we
also consider GRBs with $z<5$ and $L_{\rm iso}>L_{\rm lim}(z=5)\approx3.1\times10^{51}$ erg
$\rm s^{-1}$ (consisting of 104 GRBs). These data are delimited by the green
shaded region in Fig.~2. Compared to the sub-sample with $z<4$ and
$L_{\rm iso}>1.8\times10^{51}$ erg $\rm s^{-1}$, this new sub-sample has 12
more high-$z$ ($4<z<5$) bursts. For the cumulative redshift distribution of
these 104 GRBs between $z=0$ and $z=5$, the width of $z$ bin $\Delta z=0.125$
is chosen to ensure the number of $z$ bins (i.e., $n=40$) is the same as
that of the sub-sample with $z<4$ and $L_{\rm iso}>1.8\times10^{51}$ erg $\rm s^{-1}$.
Using the $\chi^{2}$ statistic, the constraints on the $M_{\rm min}$-$\alpha$ plane
from these 104 GRBs are shown in Fig.~5(b). It is found that adding 12 high-\emph{z}
GRBs could result in much tighter constraints on $M_{\rm min}$. The contours show that
models with $\log M_{\rm min}<7.7$ and $>11.6$ are ruled out at the $1\sigma$ confidence level.
These are in agreement with what are found by \citet{Munoz11}, in which the minimum
halo masses of $\log M_{\rm min}<8.5$ and $>9.7$ are ruled out at the $95\%$ confidence level.
At the $1\sigma$ level, the value of $\alpha$ lies in the range $0.10<\alpha<2.55$.
The cross indicates the best-fit pair $(\log M_{\rm min},\;\alpha)=(10.5,\;1.25)$.

In sum, we find that the redshift distributions of GRBs are consistent with only
moderate evolution of $(1+z)^{\alpha}$ over both $0<z<4$ and $0<z<5$ ($\sim1\sigma$ confidence).
Compared to previous studies \citep[e.g.,][]{Kistler09} the results are consistent at the
$1\sigma$ level, but we obtain a weaker redshift dependence (i.e., weaker enhancement of
the GRB rate compared to the CSFR) with lower values of $M_{\rm min}$.
In addition, the comparison between Fig.~5(a) and Fig.~5(b) may also be summarized as follows:
the best-fit results are very different for the two redshift distributions, the distribution of 104
GRBs with $z<5$ and $L_{\rm iso}>3.1\times10^{51}$ erg $\rm s^{-1}$ (see Fig.~5b) requires a relatively stronger
redshift dependence and a higher value of $M_{\rm min}$ owing to the increased number of high-\emph{z}
GRBs at $4<z<5$. Of course, there is also still a lot of uncertainty because of the small high-\emph{z} GRB sample effect.

\begin{figure}[hp]
\vskip-0.2in
\centerline{\includegraphics[angle=0,scale=1.2]{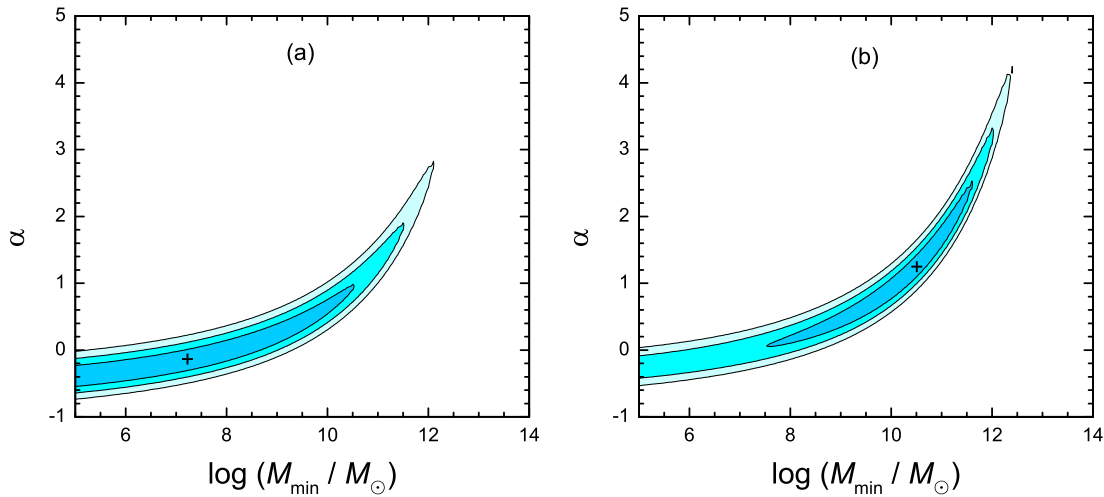}}
\vskip-0.1in
\caption{(a): $1\sigma-3\sigma$ constraint contours for $M_{\rm min}$ and $\alpha$,
    inferred from the cumulative redshift distribution of 118 \emph{Swift} GRBs with $z<4$ and
    $L_{\rm iso}>1.8\times10^{51}$ erg $\rm s^{-1}$. The cross indicates the best-fit pair
    $(\log M_{\rm min},\;\alpha)=(7.2,\;-0.15)$. (b): same as panel~(a), but for 104 \emph{Swift}
    GRBs with $z<5$ and $L_{\rm iso}>3.1\times10^{51}$ erg $\rm s^{-1}$. The cross indicates
    the best-fit pair $(\log M_{\rm min},\;\alpha)=(10.5,\;1.25)$.}\label{contour}
\end{figure}

\section{Future constraints}
\label{sect:future}

The results of our analyses suggest that the current \emph{Swift} GRB observations can in fact
be used to place some constraints on the minimum dark matter halo mass. We obtain constraints of
$7.7<\log M_{\rm min}<11.6$ at the $1\sigma$ confidence level. However, these constraints are not strong, and
they have uncertainties because of the small GRB sample effect. To increase the significance
of the constraints, one needs a larger sample. In order to investigate how much the constraints
could be improved with a larger sample, we perform some Monte Carlo simulations based on the future
mission, the Sino-French spacebased multiband astronomical variable objects monitor (\emph{SVOM}).
The \emph{SVOM} has been designed to optimize the synergy between space and ground instruments,
so it is forecast to observe $\sim 70-90$ GRBs $\rm yr^{-1}$ \citep[see, e.g.,][]{Salvaterra08}.
We simulate a sample of 450 GRBs, each of which is characterized by a set of parameters denoted as
($z$, $L_{\rm iso}$). The sample size represents an optimistic prediction of 5 yr observations of \emph{SVOM}
\citep[see, e.g.,][]{Salvaterra08,de Souza13}. The soft gamma-ray telescope ECLAIRs onboard
the \emph{SVOM} mission will provides fast and accurate GRB triggers to other onboard telescopes,
as well as to ground-based follow-up telescopes. Thanks to a low energy threshold of 4 keV,
ECLAIRs will be as sensitive as the \emph{Swift}/BAT for the detection of GRBs \citep{Godet14}.
Therefore, we adopt the same bolometric energy flux threshold of \emph{Swift},
$F_{\rm lim}=1.2\times10^{-8}$ erg $\rm cm^{-2}$ $\rm s^{-1}$, for \emph{SVOM}.
Our detailed simulation procedures are described as follows:

\begin{figure}[hp]
\vskip-0.5in
\centerline{\includegraphics[angle=0,scale=1.15]{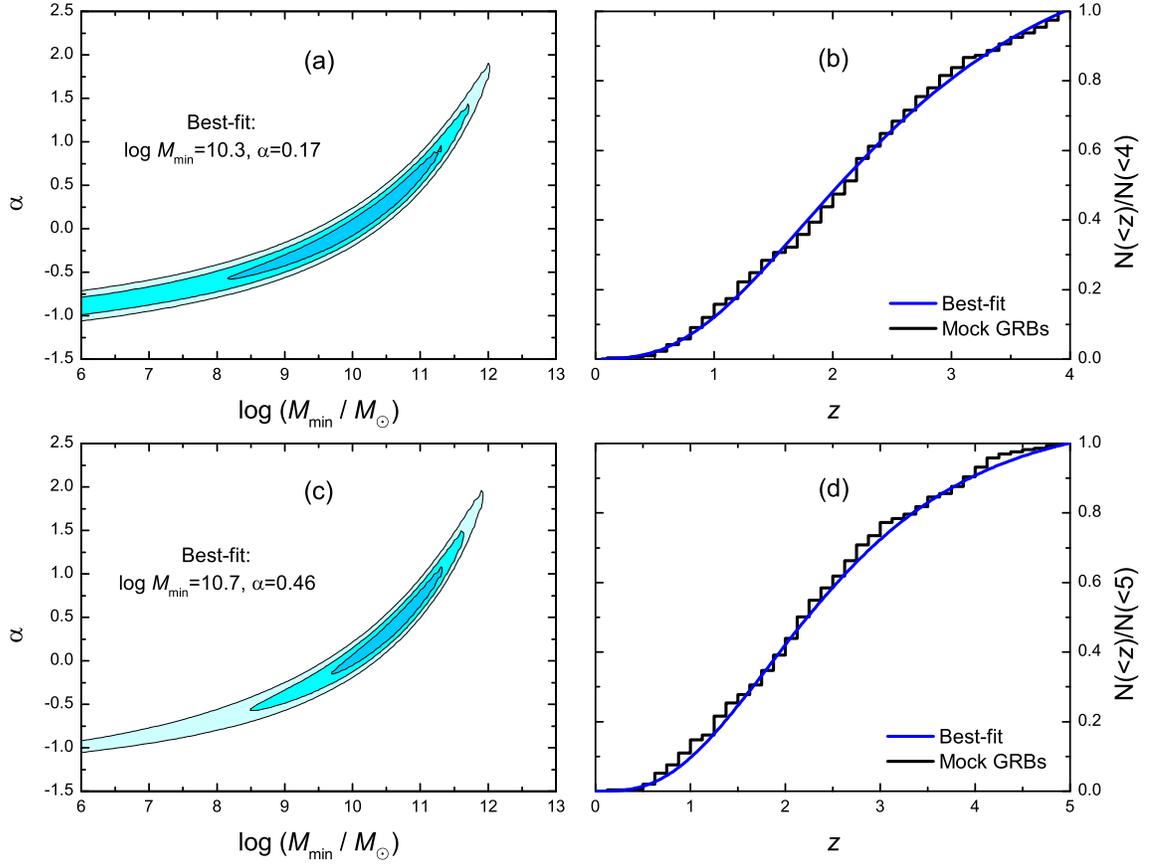}}
\vskip-0.5in
\caption{Same as Figure~5, except now for 450 mock GRBs. Left-hand panels: fitted values of
$M_{\rm min}$ and $\alpha$ using the cumulative redshift distribution of 310 mock GRBs
with $z<4$ and $L_{\rm iso}>1.8\times10^{51}$ erg $\rm s^{-1}$ or the distribution of
291 mock GRBs with $z<5$ and $L_{\rm iso}>3.1\times10^{51}$ erg $\rm s^{-1}$ (right-hand
panels; steps). The theoretical curves (right-hand panels; solid lines) correspond to
the parameter values that minimize the $\chi^{2}$ (shown on the left).}\label{mock}
\end{figure}

1. The redshift $z$ is generated randomly from the co-moving number density of GRBs
at redshift $\rm \emph{z}+d\emph{z}$, i.e., $\Re(z)=\frac{\dot{n}_{\rm GRB}(z)}{1+z}\frac{\rm d \emph{V}_{com}}{\rm d\emph{z}}$.
We consider that the GRB rate follows the CSFR, $\dot{n}_{\rm GRB}(z)\propto\dot{\rho}_{\star}(z)$.
For the CSFR $\dot{\rho}_{\star}(z)$, we adopt the empirical fit model \citep{Hopkins06,Li08}
\begin{equation}\label{SFRone}
\log \dot{\rho}_{\star}(z)=a+b\log_{10}(1+z)\;,
\end{equation}
where
\begin{equation}
  (a,b) = \left\lbrace \begin{array}{lll}(-1.70, 3.30), ~~~~~~~~~~z <0.993\\
                                         (-0.727, 0.0549), ~~~~~0.993< z <3.8\;. \\
                                         (2.35, -4.46), ~~~~~~~~~~ z >3.8 \\
\end{array} \right.
\end{equation}
Since $z<10$ for the current $Swift$ sample, the range of $z$ for our analysis is from 0 to 10.

2. The intrinsic luminosity distribution for LGRBs has been well constrained by \citet{Wanderman10},
which is a simple broken power law function,
\begin{equation}
  \Phi(L) = \left\lbrace \begin{array}{ll}\left(L/L_{\star}\right)^{x}, ~~~~~~~L <L_{\star}\;, \\
                                          \left(L/L_{\star}\right)^{y}, ~~~~~~~L >L_{\star}\;, \\
\end{array} \right.
\end{equation}
where $x=-0.65$, $y=-3$, and $L_{\star}=10^{52.05}$ erg $\rm s^{-1}$.
The mock luminosity $L_{\rm iso}$ is obtained by sampling the probability density function given by Equation~(15).

3. With the mock $z$ and $L_{\rm iso}$, the bolometric energy flux is calculated by $F=L_{\rm iso}/4\pi D_{L}^{2}(z)$.
If $F>F_{\rm lim}$, a mock GRB is recognized as detectable. Otherwise, the mock GRB is excluded.

4. Repeat the above steps to obtain a sample of 450 GRBs.

As described above, we only choose the luminous mock bursts for analysis to reduce the selection effects
and also consider two sub-samples (i.e., (S1) the sub-sample with $z<4$ and $L_{\rm iso}>1.8\times10^{51}$
erg $\rm s^{-1}$ and (S2) the sub-sample with $z<5$ and $L_{\rm iso}>3.1\times10^{51}$ erg $\rm s^{-1}$)
to explore the dependence of our results on a possible bias in the high-\emph{z} bursts.
Using the $\chi^{2}$ statistic, the constraints on the $M_{\rm min}$-$\alpha$ plane from the cumulative redshift distribution
of the S1 sub-sample are presented in Fig.~6(a). These contours show that at the $1\sigma$ confidence level,
we have $8.2<\log M_{\rm min}<11.3$, and $-0.57<\alpha<0.95$. The constraints on these
two parameters from the S2 sub-sample are also shown in Fig.~6(c). As the results we find in the current $Swift$
GRB observations, adding the high-\emph{z} $(4<z<5)$ mock GRBs could result in much tighter constraints
on $M_{\rm min}$ and relatively higher values of $\alpha$ and $M_{\rm min}$.
The contours show that $\sim5$ yr mission of $SVOM$ (using the S2 sub-sample) would
be sufficient to rule out $\log M_{\rm min}<9.7$ and $>11.3$ models at the $1\sigma$ confidence level.
The evolutionary index is constrained to be $-0.11<\alpha<1.08$ ($1\sigma$).
From these results, it is evident that as the sample size increases, the constraints on $M_{\rm min}$ and $\alpha$
become tighter than the current constraints using the $Swift$ sample.

\section{Discussion and Conclusions}
\label{sect:Conclusion}
Using the hierarchical structure formation scenario, the CSFR can be built in a self-consistent way.
In particular, from the hierarchical scenario we can obtain the baryon accretion rate that governs
the size of the reservoir of baryons available for star formation in dark matter halos. It is important
to note that the minimum halo mass $M_{\rm min}$ plays an important role in star formation, because
first stars can only form in structures that are suitably dense. Star formation will be
suppressed when the halo mass is below $M_{\rm min}$. The connection of LGRBs with the collapse of
massive stars has provided a good opportunity for probing star formation in dark matter halos.

In this paper, the numerical value of $M_{\rm min}$ is constrained using the latest $Swift$ GRB data.
We conservatively consider that the LGRB rate is proportional to the CSFR and an additional evolution
parametrized as $(1+z)^{\alpha}$. In order to reduce the sample selection effects, we adopt a
model-independent approach by selecting only luminous GRBs above a fixed luminosity limit, as
\citet{Kistler08} did in their treatment. This approach has two advantages. Firstly, the reliable
statistics of the latest LGRB data allow the use of luminosity cuts to fairly compare GRBs in the full redshift
range, eliminating the unknown GRB luminosity function. Secondly, by simply normalizing the cumulative
redshift distribution of GRBs to the full redshift range, the constant stands for the GRB efficiency factor
can be removed.

For each model ($M_{\rm min}$, $\alpha$), we can calculate the expected cumulative redshift distribution.
The confidence limits in the $M_{\rm min}$-$\alpha$ plane can be constructed by fitting the observed
cumulative redshift distribution, using the $\chi^{2}$ statistic.
Our results show that at the $1\sigma$ confidence level, we obtain $M_{\rm min}<10^{10.5}$
$\rm M_{\odot}$ from 118 \emph{Swift} GRBs with $z<4$ and
$L_{\rm iso}>1.8\times10^{51}$ erg $\rm s^{-1}$. We also find that adding 12 high-\emph{z} $(4<z<5)$
GRBs (comprised of 104 GRBs with $z<5$ and $L_{\rm iso}>3.1\times10^{51}$ erg $\rm s^{-1}$)
could result in much tighter constraints on $M_{\rm min}$, for which, $10^{7.7}\rm M_{\odot}<M_{\rm min}<10^{11.6}\rm M_{\odot}$
at the $1\sigma$ confidence level. Through Monte Carlo simulations, we find that
the constraints on $M_{\rm min}$ and $\alpha$ can be much improved by enlarging the sample size.
The simulations show that the future $SVOM$ 5-yr observations would tighten these constraints to
$10^{9.7}\rm M_{\odot}<M_{\rm min}<10^{11.3}\rm M_{\odot}$ at the $1\sigma$ confidence level.

Previously, with a minimum halo mass of $10^{7}-10^{8}M_{\odot}$ and a moderate outflow
efficiency, \citet{Daigne06b} could reproduce both the fraction of baryons in the structures
at the present time and the early chemical enrichment of the intergalactic medium.
By analysing the star formation history, \citet{Bouche10} set a strong constraint
on the minimum halo mass: $M_{\rm min}\simeq10^{11}M_{\odot}$. \citet{Munoz11} also suggested that
the halo mass at which star formation is suppressed can be limited by matching the observed
galaxy luminosity distribution, in which $M_{\rm min}$ was constrained to be
$10^{8.5}\rm M_{\odot}<M_{\rm min}<10^{9.7}\rm M_{\odot}$ at the $95\%$ confidence level.
In the present paper, we propose that $M_{\rm min}$ can also be constrained
using the redshift distribution of \emph{Swift} GRBs, and we obtain some limits on $M_{\rm min}$,
namely $10^{7.7}\rm M_{\odot}<M_{\rm min}<10^{11.6}\rm M_{\odot}$ ($1\sigma$), which
are consistent with the previous results obtained using both the current baryon fraction
and the early chemical enrichment of the intergalactic medium, the star formation history,
and the galaxy luminosity function. Although the future $SVOM$ 5-yr observations would
tighten these constraints to $10^{9.7}\rm M_{\odot}<M_{\rm min}<10^{11.3}\rm M_{\odot}$
($1\sigma$), the lower limit value of $M_{\rm min}$ is above the upper limit given by
\citet{Munoz11}, and well above the values of \citet{Daigne06b}.

The strong constraints we derived here indicate that LGRBs are a new promising tool
for probing star formation in dark matter halos. Of course, if we know the mechanism responsible
for the difference between the LGRB rate and the CSFR, we can constrain the minimum mass very accurately
using the LGRB data alone and the utility of LGRBs would be further enhanced.
Apart from the obvious approach of increasing the sample size of LGRBs in the future, we predict that the constraints on $M_{\rm min}$
will also be significantly improved by including different types of observational data, such as the data of
star formation history, galaxy luminosity distribution, and GRB redshift distribution.

\section*{Acknowledgments}
We acknowledge the anonymous referee for his/her important suggestions,
which have greatly improved the manuscript. We also thank Z. G. Dai, Y. F. Huang,
X. Y. Wang, F. Y. Wang, and W. W. Tan for helpful discussions.
This work is partially supported by the National Basic Research Program (``973" Program)
of China (Grants 2014CB845800 and 2013CB834900), the National Natural Science Foundation
of China (grants Nos. 11073020, 10733010, 11133005, 11322328, and 11433009),
the One-Hundred-Talents Program, the Youth Innovation Promotion Association (2011231), and
the Strategic Priority Research Program ``The Emergence of Cosmological Structures"
(Grant No. XDB09000000) of the Chinese Academy of Sciences.

%

%
\end{document}